\magnification=\magstephalf
\font\t=cmcsc10 at 13 pt
\font\tt=cmcsc10
\font\n=cmcsc10
\font\foot=cmr9
\font\foots=cmsl9
\font\abs=cmr8
\font\babs=cmbx8
\font\sm=cmsl8
\centerline{\t Mechanics. - On the analytic expression that must be given to}
\smallskip
\centerline{\t the gravitational tensor in Einstein's theory \footnote{\dag}
{\foot{Rendiconti della Reale Accademia dei Lincei
{\bf 26}, 381 (1917).}}}\bigskip
\centerline{Note by the Fellow}\bigskip
\centerline{\n T. Levi-Civita}\bigskip
\vbox to 0.6 cm {}
\centerline{\tt translation and foreword by}\smallskip
\centerline{S. Antoci\footnote{$^*$}{\foot Dipartimento di Fisica
``A. Volta'', Universit\`a di Pavia, Via Bassi 6 - 27100 Pavia
(Italy).} and A. Loinger\footnote{$^{**}$}{\foot Dipartimento di Fisica,
Universit\`a di Milano, Via Celoria 16 - 20133 Milano
(Italy).}}\medskip
{\babs Foreword.} {\abs While most textbooks of general relativity and
research articles discuss at length the relative merits of the
pseudo tensors proposed by Einstein and by other authors for representing
the energy of the gravitational field, Levi Civita's definition of a true
gravitational energy tensor has succumbed to Einstein's authority and is
nearly forgotten. It seems however worthy of a careful re-examination,
due to its unquestionable logical soundness and to the unique manner of
propagation for gravitational energy that it entails.}

\vbox to 0.6 cm {}
    In the present Note, after having recalled, for the reader's
convenience, the leading idea and the mathematical framework of
general relativity, I show how some identities (involving the
derivatives of the Riemann symbols) discovered by Bianchi offer a
sure criterion for introducing the so called gravitational
tensor.~From the analytic standpoint one has to do with a double symmetric
system $A_{ik}~(i,k=0,1,2,3)$, whose ten elements completely
define the gravitational contribution to the local mechanical
behaviour. In fact they determine the stresses as well as the energy
flow and the energy density (of gravitational origin). The mechanical
meaning of the system requires an analytic structure endowed with
convenient invariance properties with respect to co-ordinate
transformations. Such is actually the (covariant) form of the
$A_{ik}$ yielded by the above mentioned criterion. Furthermore
this form provides a very expressive extension of
d'Alembert's principle.\par
    The idea of a gravitational tensor belongs to the great
construction by Einstein. However its definition as given by the
Author cannot be considered final. First of all, from the
mathematical standpoint, it lacks the invariant character that it
should instead necessarily enjoy according to the spirit of general
relativity. Even worse is the fact, perceived with keen intuition
by Einstein himself \footnote{($^1$)}{\foots{N\"aherungweise
Integration der Feldgleichungen der Gravitation,}
\foot{Sitzungsberichte der Kgl. Preussischen Ak. der Wiss., 1916,
p. 696.}}, that from such a definition it follows a clearly
unacceptable consequence about the gravitational waves. For this
point he however finds a way out in quantum theory.\par
    The solution is however less remote: everything depends on
the incorrect form assumed for the gravitational tensor. We shall
see that with our determination any possibility for paradox
automatically disappears.\par\medskip
{\tt 1. Generalities.} - In ordinary mechanics the physical
space is taken to be strictly Euclidean, and the analytical
representation of phenomena is, let us say, subordinated to the
(ternary) quadratic differential form $dl^2$ that expresses the
square of the line element.\par
    In the restricted theory of relativity one persists in
considering space as Euclidean; however the equations of mechanics
are no longer invariant with respect to the form $dl^2$; they are
invariant with respect to a quaternary form $ds^2$ that implies also
the time $t$.
Notoriously it has the expression
$$ds^2=c^2dt^2-dl^2\leqno(1)$$
($c$ universal constant to be interpreted as velocity of light
{\it in vacuo}).\par
It is clear that, with reference to Cartesian co-ordinates, one
has again
$$dl^2=dx^2+dy^2+dz^2.$$

    In general relativity - new and more comprehensive conception
of the natural laws, again due to Einstein - space and time do not
provide a plain localisation, inert and immutable, of phenomena.
They are instead affected by the latter and they react in such a
way as to change the nature of $ds^2$.\par
    Instead of (1), one has the fundamental form
$$ds^2={\sum_0^3}{\sm ik}~g_{ik}dx_idx_k,\leqno(2)$$
which, with appropriate choice of the parameters $x_0$, $x_1$,
$x_2$, $x_3$, reduces itself exactly to the form (1) in the limit
case, when any physically perceptible action (either presence
or motion of matter, of electricity, more generally of some form
of energy) is lacking. As a rule, although quantitatively very
close to the type (1), (2) must be considered as not given {\it a
priori}, but intrinsically definable according to the factual
circumstances. Among these it obviously appears also the universal
gravitation, which, according to Einstein, deserves the privilege
of depending exclusively on the coefficients $g_{ik}$ (and on their
derivatives).\par
    The equations of the new mechanics as a whole are invariant with
respect to that well determined form (2) that pertains to the
specific case. In this new mechanics the theory for a given class
of phenomena necessarily entails, together with relations that
have their counterpart in the previous formulations (classical and
relativistic of the first manner), further relations whose scope
is the determination of $ds^2$. These are Einstein's gravitational
equations (in number of 10 like the coefficients $g_{ik}$), that
we shall consider explicitly in \S~6.\par\medskip

{\tt 2. Energy tensor.} - The mechanics of continuous systems -
also according to the ordinary scheme - leads to conclude that a
mechanical phenomenon (occurring in a given range of values for
$x$, $y$, $z$, $t$) is well known when the following elements:
stresses, momentum, energy flow and energy density, are assigned
(as functions of position and of time).\par

In relativistic mechanics the vector ${\bf q}$, that represents
the momentum density, is linked to the energy flow $\chi$ by the
relation

$${\bf q}={1\over c^2}\chi.$$
It is convenient to avail of the single vector

$$-{\bf f}=c{\bf q}={1\over c}\chi,\leqno(3)$$
that can be considered as the energy flow occurring in a
light-second (time interval during which light travels the length
unit). Let us notice, to avoid any misunderstanding, that we do
not mean to fix in this way the time unit: it remains generic, like
the other two fundamental units.\par
    Let us write for short

$$y_0=ct,~~y_1=x,~~y_2=y,~~y_3=z.\leqno(4)$$
With reference to these variables, let us introduce a
symmetric tensor $T_{ik}$ defined in this way: for $i,k=1,2,3$,
$T_{ik}$ is the component along the $y_k$ axis of the
specific stress exerted on a surface element normal to the $y_i$
axis \footnote {$(^2)$}{\foot with the convention (usual in
hydrodynamics) that a positive normal stress corresponds to
pressure.} (or vice versa, by exchanging $i$ and $k$);
$T_{i0}=T_{0i}$ is identified with the component $f_i$ of the
vector ${\bf f}$; $T_{00}$ is finally the energy
density.\par\medskip
{\tt 3. Reference to arbitrary co-ordinates.} - If four arbitrary
(independent) combinations $x_0$, $x_1$, $x_2$, $x_3$ are
substituted for the $y_i$, the form
$$ds^2=c^2dy_0^2-(dy_1^2+dy_2^2+dy_3^2)\leqno(1')$$
takes the general expression (2). However the qualitative
restrictions
$$g_{00}>0,~~g_{ii}<0~~~~(i=1,2,3),\leqno(5)$$
shall hold whenever the parameter $x_0$ (individually varied) is
apt to reflect the intuitive notion of time, while the remaining
$x_i$ can in some way be interpreted as actual space
co-ordinates.\par
    When this holds, {\it we shall interpret as energy tensor in
arbitrary co-ordinates $x_i$ that double covariant system
$T_{ik}$ which, when referred to the $y$, is specified in the way
shown above.}\par
    From the very covariance formulae that define the elements of
the system $T_{ik}$ one derives, for each of them, an
interpretation in general co-ordinates $x$. And precisely one
finds \footnote{$(^3)$} {\foot Dwelling here on the way of
deduction would be out of place. However I allow myself to notice that
there is no need of actual calculations: one can avail of an
appropriate adjustment of the methods of the absolute differential
calculus to the indefinite $ds^2$ that encompass space and time.}
that:\par

$${T_{ik}\over\sqrt{g_{ii}g_{kk}}}
={T_{ki}\over\sqrt{g_{ii}g_{kk}}}~~~(i,k=1,2,3)$$
represents the (orthogonal) component along the line $x_i$ (the
one along which only $x_i$ varies) of the stress exerted on a
surface element normal to the line $x_k$ (or vice versa,
by exchanging $i$ and $k$);

$${T_{i0}\over\sqrt{-g_{00}g_{ii}}}
={T_{0i}\over\sqrt{-g_{00}g_{ii}}}~~~(i=1,2,3)$$
represents the component of ${\bf f}$ along the line $x_i$ (when
one imagines decomposing the vector ${\bf f}$ with respect to
the trihedron of the co-ordinate lines); finally

$$T_{00}\over g_{00}$$
is the density of the energy distribution in the space
$(x_1,~x_2,~x_3)$, [to which the metric determination
corresponding to $-ds^2$ for $dx_0=0$ is attributed].\par\medskip
    {\tt 4. Linear invariant and divergence of the energy tensor.} -
We adhere to the usual notation of the absolute differential calculus.
Therefore we represent with $g^{(ik)}$ the elements reciprocal to
the coefficients $g_{ik}$, and with $T_{ikl}~(i,k,l=0,1,2,3)$ the
system covariantly derived from $T_{ik}$ according to the
fundamental form. At present, as already in the previous $\S$, we
shall assume this form to be (1') which, when referred to arbitrary
co-ordinates $x$, takes the generic expression (2).\par
    By setting
$$T={\sum_0^3}{\sm ik}~g^{(ik)}T_{ik},\leqno(6)$$
one defines an invariant, that is just called {\it linear
invariant} or {\it scalar of the energy tensor}.\par
    One calls instead {\it divergence} of the same energy tensor
the simple covariant system (or four-dimensional vector)

$$F_i={\sum_0^3}{\sm kl}~g^{(kl)}T_{ikl}~~~(i=0,1,2,3).\leqno(7)$$

The mechanical meaning of the divergence (like the meaning of $T$,
which I neglect to notice, because it is immediate) becomes clear
when one goes back to the variables $y$. With respect to such
variables $g^{(ik)}=0~(i\neq k)$, $g^{(00)}=1$,
$g^{(ii)}=-1~(i=1,2,3)$, and the covariant differentiation
coincides with the ordinary one.\par
    Then one has
$$\eqalign{F_i&={\partial T_{i0}\over\partial y_0}
-{\sum_1^3}{\sm k}~{\partial T_{ik}\over\partial y_k}~~~(i=1,2,3),\cr
F_0&={\partial T_{00}\over\partial y_0}
-{\sum_1^3}{\sm k}~{\partial T_{0k}\over\partial y_k}.}$$

    Let us remind of $\S~2$ and notice that, due to (3), the
components $T_{i0}$ are identical with $-cq_i$ ($q_i$ components of
the momentum density ${\bf q})$. Therefore it obviously appears
from the first three equations written above (when $ct$ is
substituted for $y_0$) that $-F_i~(i=1,2,3)$ are components of the
external force ${\bf F}$ applied to the system (per unit volume).
The last equation, when [always according to (3)] the $T_{0k}$ are
taken in the form $-{1\over c}\chi_k$ ($\chi_k$ components of the
energy flow $\chi$), finally shows that $cF_0$ is the power
density, {\it i.e.} the energy given from outside to the system
per time and volume units. One can also say, if desired, that
$F_0$ represents the energy given to the system per volume unit in
a light-second. Hence it turns out in particular that for an
isolated system the divergence is vanishing.\par
    When one avails as reference of arbitrary co-ordinates $x_i$,
the covariant character of the simple system $F_i$ immediately
allows one to interpret

$${-F_1\over\sqrt{-g_{11}}},~~{-F_2\over\sqrt{-g_{22}}},
~~{-F_3\over\sqrt{-g_{33}}}$$
as components of ${\bf F}$ along the co-ordinate lines $x_1$,
$x_2$, $x_3$;

$$F_0\over\sqrt{g_{00}}$$
as energy given in one light-second to the unit volume of the
system.\par\medskip
    {\tt 5. Transition to general relativity.} - Despite the
reference to general co-ordinates, up to now we have supposed to
deal with a Euclidean $ds^2$. Formally the situation happens to
be the same also with a $ds^2$ essentially not reducible to the
type (1'), if however:\par
    a) $x_0$ can be interpreted as time and the other three
co-ordinates as space parameters since, in keeping with this,
the inequalities (5) hold true;\par
    b) the usual mechanical intuitions are preserved (at an
infinitesimal scale), hence it is possible to attribute a definite
meaning to local measurements of force, stresses, energy flow and
energy density. In such conditions the energy tensor is uniquely
defined through the ratios
$${T_{ki}\over\sqrt{g_{ii}g_{kk}}}~~~(i,k=1,2,3)~;
~~{T_{0i}\over\sqrt{-g_{00}g_{ii}}}~~~(i=1,2,3)~;
~~{T_{00}\over g_{00}},$$
given in $\S~3$.\par
    We shall hereafter assume our $ds^2$ to be {\it a priori}
arbitrary (apart from the restrictions given above); naturally
one shall take this $ds^2$ as fundamental form.\par\medskip
    {\tt 6. The equations of the gravitational field.} - Let
$g_{ij,hk}~(i,j,h,k=0,1,2,3)$ indicate the Riemann symbols of the
first kind belonging to a generic quaternary $ds^2$ like (2).
Due to their covariance, the positions
$$G_{ik}={\sum_0^3}{\sm jh}~g^{(jh)}g_{ij,hk}~~~
(i,k=0,1,2,3)\leqno(8)$$
define a double covariant system.\par
    Let us remind of the formulae \footnote{$(^4)$}{{\foot
Bianchi,} {\foots Lezioni di geometria differenziale}, {\foot vol.
I [Pisa, Spoerri, 1902], p. 72.}}

$$g_{ij,hk}={\sum_0^3}\nu~g_{j\nu}\{i\nu,hk\},$$
$$\{i\nu,hk\}={\partial\over{\partial x_k}}\Big\{^{i~h}_{~\nu}\Big\}
-{\partial\over{\partial x_h}}\Big\{^{i~k}_{~\nu}\Big\}
+{\sum_0^3}{\sm l}~
\Big[\Big\{^{i~h}_{~l}\Big\}\Big\{^{l~k}_{~\nu}\Big\}
-\Big\{^{i~k}_{~l}\Big\}\Big\{^{l~h}_{~\nu}\Big\}\Big],$$
that connect the Riemannian symbols of the first kind with those
of the second kind, and the latter with the Christoffel
symbols (again of the second kind). One immediately recognises
that (8) are equivalent to

$$\eqalign{G_{ik}&={\sum_0^3}{\sm h}~\{ih,hk\}=\cr
&={\sum_0^3}{\sm h}~\Big[{\partial\over{\partial x_k}}\Big\{^{i~h}_{~h}\Big\}
-{\partial\over{\partial x_h}}\Big\{^{i~k}_{~h}\Big\}\Big]
+{\sum_0^3}{\sm hl}~
\Big[\Big\{^{i~h}_{~l}\Big\}\Big\{^{k~l}_{~h}\Big\}
-\Big\{^{i~k}_{~l}\Big\}\Big\{^{l~h}_{~h}\Big\}\Big].}\leqno(8')$$
The linear invariant of the double system $G_{ik}$

$$G={\sum_0^3}{\sm ik}~g^{(ik)}G_{ik}\leqno(9)$$
will be called {\it average curvature} of our $ds^2$ \footnote {$(^5)$}
{\foot This name is obviously derived from the geometric meaning that
$G$ would take, if $ds^2$ would be positive definite.}.
    With these positions, Einstein's gravitational equations are:

$$G_{ik}-{1\over 2}g_{ik}G=-\kappa T_{ik},\leqno(10)$$
where $\kappa$ depends on the constant $f$ of universal
gravitation and on $c$ according to the formula

$$\kappa={8\pi f\over c^4}.\leqno(11)$$
I remark in passing that the homogeneity of both sides of
(10) can be checked if one imagines referring to mutually
homogeneous parameters $x_0$, $x_1$, $x_2$, $x_3$, {\it e.g.}
lengths, like (for the Euclidean $ds^2$) the $y$ defined by the
positions (4). The coefficients $g_{ik}$ are then pure numbers,
and the left-hand sides have clearly the dimensions $l^{-2}$. On
the other hand all the $T_{ik}$ (specific stresses apart from
numerical factors, etc.) have in this case the same
dimensions, and precisely $ml^{-1}t^{-2}$. One has further

$$[f]=m^{-1}l^3t^{-2}~~,~~[\kappa]=m^{-1}l^{-1}t^2,$$
hence also the right-hand sides have actually the dimensions
$l^{-2}$.\par\medskip

    {\tt 7. Formal validation derived from the Bianchi identities.} -
The covariant derivatives of the Riemann symbols are linked by
very remarkable relations due to Bianchi \footnote {$(^6)$}
{\foot see loc. cit., p. 351.}, that can be resumed in the formula

$$g_{ij,hkl}+g_{jl,hki}+g_{li,hkj}=0~~~(i,j,h,k,l=0,1,2,3),$$
or, due to well known properties of the Riemann symbols, in
the equivalent formula

$$g_{ij,hkl}+g_{il,khj}-g_{lj,hki}=0.$$

    Let us multiply by ${1\over 2}g^{(kl)}g^{(jh)}$ and sum with
respect to $k$, $l$, $j$, $h$; when the sum is accomplished, let
us exchange in the second term $j$ with $l$ and $h$
with $k$. The second term thus becomes identical with the first,
and one gets

$${\sum_0^3}{\sm kljh}~g^{(kl)}g^{(jh)}g_{ij,hkl}
-{1\over 2}{\sum_0^3}{\sm kljh}~
g^{(kl)}g^{(jh)}g_{lj,hki}=0~~~(i=0,1,2,3).$$
On the other hand, the covariant differentiation of (8) - by
reminding of Ricci's lemma, according to which the coefficients of
the fundamental form have vanishing covariant derivative - yields

$$G_{ikl}={\sum_0^3}{\sm jh}~g^{(jh)}g_{ij,hkl}.$$

\noindent From the expression (9) of $G$, that can be written as

$$G={\sum_0^3}{\sm kl}~g^{(kl)}G_{lk},$$
by covariant differentiation one gets

$${\partial G\over{\partial x_i}}=G_i
={\sum_0^3}{\sm kl}~g^{(kl)}G_{lki}
={\sum_0^3}{\sm kljh}~g^{(kl)}g^{(jh)}g_{lj,hki}.$$
The resulting combinations of the Bianchi identities thus become

$${\sum_0^3}{\sm kl}~g^{(kl)}G_{ikl}-{1\over 2}G_i=0~~~
(i=0,1,2,3).\leqno(12)$$
They contain the validation of the gravitational equations (10)
from the mathematical standpoint. Here is why: the right-hand
sides of (10) constitute a double system with vanishing divergence
\footnote {$(^7)$} {\foot In fact $T_{ik}$ include the
contribution of all the phenomena that occur at the considered
place and time (apart from gravitation). One deals anyway with an
isolated system in the ordinary sense of the word. Therefore force
and power must vanish within each of its elementary portions.}. If
one requires the system (10) to be complete [{\it i.e.} no
condition is imposed on $ds^2$ beyond the exterior circumstances
resumed in $T_{ik}$], also the divergence of the first-hand sides,
hence of the system

$$G_{ik}-{1\over 2}g_{ik}G,$$
must identically vanish. This fact is just expressed by the
equations (12).\par\medskip
    {\tt 8. Gravitational (or inertial) tensor. - Generalisation
of d'Alembert's principle.} - If we set for short

$$A_{ik}={1\over \kappa}
\bigg\{G_{ik}-{1\over 2}g_{ik}G\bigg\},\leqno(13)$$
the gravitational equations (10) read

$$T_{ik}+A_{ik}=0~~~(i,k=0,1,2,3).\leqno(10')$$

    In them we interpret $A_{ik}$ as components of an energy tensor
due to the space-time environment, {\it i.e.} exclusively
dependent on the coefficients of $ds^2$. Such a tensor can be
equally well named either {\it gravitational} or {\it inertial}
\footnote {$(^8)$} {{\foot See for instance W. De Sitter,}
{\foots On the relativity of rotation in Einstein's theory},
{\foot Proc. of K. Ak. van Wet. te Amsterdam, vol. XIX, 1916,
p. 530 (in footnote).}} because both gravitation and inertia
depend on $ds^2$. Therefore (10') give rise to the following
proposition:\par
    {\it The nature of $ds^2$ is always such as to balance all
mechanical actions; in fact the sum of the energy tensor and of
the inertial one identically vanishes.}\par
    One is naturally led to associate this proposition with
d'Alembert's principle ``the lost forces ({\it i.e.} directly
applied forces and inertial ones) balance each other''. The
equilibrium expressed by (10') is just the most complete
occurrence that can be conceived from the mechanical standpoint.
In fact, not only the total force applied to each single element
comes to vanish, but also stresses, energy flow and energy density
(by taking inertia into account through $A_{ik}$)
behave in this way.\par
    It is clear that this total lack of mechanical entities
pertains to isolated systems. Let us introduce in the field of
such a system for instance a bit of matter (and for simplicity the
ensuing alteration of the field is supposed negligible); several
external actions coming from the system are exerted
on the extra matter. In the ideal case of the mass point,
these can be summarised in a law of motion (geodesic with respect to the
four-dimensional $ds^2$). It contains in particular the ordinary
dynamics of a point subjected to conservative forces.\par
    It must be remarked that Einstein's fundamental equations,
connected here with d'Alembert's principle, have been already
derived, by Einstein himself and, in a more complete way, by
Lorentz and by Hilbert \footnote{$(^9)$} {{\foot See for instance
pp. 707-709 of De Sitter's report,} {\foots On Einstein's theory
of gravitation...,} {\foot Monthly Notices, vol. LXXVI, 1916.}},
from the appropriate variation of a unique integral. In this way also
Hamilton's principle is extended to the new mechanics.\par\medskip
    {\tt 9. Einstein's misunderstanding about the gravitational
tensor.} - I remember, although it may be superfluous, that by setting

$$A_i^{(j)}={\sum_0^3}{\sm k}~g^{(jk)}A_{ik}~~~(i,j=0,1,2,3),$$
from any double covariant system $A_{ik}$ one immediately gets a mixed
system $A_i^{(j)}$ (covariant with respect to the index $i$ and
contravariant with respect to the index $j$). It follows that, in
order to identify the gravitational tensor with respect to certain
variables, it makes no difference if one fixes either the elements
$A_{ik}$ or their linear combinations $A_i^{(j)}$. With this
proviso let us come to the explicit expressions proposed by
Einstein \footnote {$(^{10})$} {{\foot Firstly with reference
to special variables, then by extending their validity; eventually
by attributing to them a general character. See in particular the
recent Note:} {\foots Hamiltonsches Prinzip und allgemeine
Relativit\"atstheorie,} {\foot Sitzungsberichte der Kgl.
Preussischen Ak. der Wiss., 1916, pp. 1111-1116.}} for the
$A_i^{(j)}$, and by him called ${\sqrt{-g}}~t_i^j$
($g$ is the discriminant of $ds^2$).\par
    They are

$${\sqrt{-g}}~t_i^j
={1\over 2}\Bigg\{G^*\varepsilon_i^j-{\sum_0^3}'{\sm hk}~
{\partial G^*\over
{\partial g_j^{(hk)}}}g_i^{(hk)}\Bigg\}~~~
(i,j=0,1,2,3).\leqno(14)$$
Here $\varepsilon_i^j$ is as usual either zero or one according to
whether the indices are different or equal; $g_j^{(hk)}$ stands for
${\partial g^{(hk)}/\partial x_j}$; the summation ${\sum_0^3}'{\sm hk}$
must be extended to all the combinations with repetition of the
indices $h$ and $k$; finally the function

$$G^*=-{\sum_0^3}{\sm ik}~g^{(ik)}{\sum_0^3}{\sm hl}~
\Big[\Big\{^{i~h}_{~l}\Big\}\Big\{^{k~l}_{~h}\Big\}
-\Big\{^{i~k}_{~l}\Big\}\Big\{^{l~h}
_{~h}\Big\}\Big].\leqno(15)$$
must be understood (as it is obviously allowed) as reduced to
depend only on the arguments $g^{(hk)}$, $g^{(hk)}_j$ before being
subjected to partial differentiation with respect to the
latter.\par
    The inappropriateness of the positions (14) from the
mathematical standpoint is easily acknowledged. It is sufficient
for instance to derive from them the expression that should
be assumed by the linear invariant, {\it i.e.}

$${\sqrt{-g}}~{\sum_0^3}{\sm i}~t_i^i
={1\over 2}\Bigg\{4G^*-{\sum_0^3}{\sm i}~{\sum_0^3}'{\sm hk}~
{\partial G^*\over
{\partial g_i^{(hk)}}}g_i^{(hk)}\Bigg\}.$$
Since $G^*$, according to (15), is quadratic and homogeneous with
respect to the Christoffel symbols, hence also with respect to
$g_i^{(hk)}$, by virtue of Euler's theorem

$${\sum_0^3}{\sm i}~{\sum_0^3}'{\sm hk}~{\partial G^*\over
{\partial g_i^{(hk)}}}g_i^{(hk)}=2G^*,$$
and {\it the invariant in question should reduce itself to $G^*$}.\par
    Now it is well known \footnote {$(^{11})$} {{\foot See for
instance Ricci et Levi-Civita,} {\foots M\'ethodes de calcul
diff\'erentiel absolu et leurs applications,} {\foot
Matematische Annalen, B. 54, 1900, p. 162.}} that
differential invariants of the $1^0$ order which are intrinsic
{\it i.e.}, like $G^*$, exclusively formed with the coefficients
of $ds^2$ and with their first derivatives, do {\it not} exist.
This is enough to render, at least in general, not admissible
the form of the gravitational tensor taken by Einstein. The latter
however had already felt some uneasiness, in particular when \footnote
{$({^{12}})$} {\foot In the Note already cited at the beginning.},
after having outlined with genial simplicity the theory of the
gravitational waves, he was led to the unacceptable result that also
{\it spontaneous} waves should as a rule give rise to dispersion of
energy through irradiation.\par
    ``Since this fact'' - these are his words - ``should not happen in
nature, it seems likely that quantum theory should intervene
by modifying not only Maxwell's electrodynamics, but also the new
theory of gravitation''.\par
    Actually there is no need of reaching to quanta. It is enough to
correct the formal expression of the gravitational tensor in the
way shown here. Then the possibility of being confronted with
consequences not corresponding to the physical intuition is {\it a
priori} excluded, in the case either of free waves or of another
{\it purely} gravitational phenomenon. In fact, by virtue of (10')
or, if one likes, of the generalised d'Alembert's
principle, when the energy tensor $T_{ik}$ vanishes, the same
occurrence must happen to the gravitational tensor $A_{ik}$. This
fact entails total lack of stresses, of energy flow, and also of a
simple localisation of energy.
\end